\documentstyle[12pt,epsf]{article}

\renewcommand{\bar}[1]{\overline{#1}}

\newcommand{\VEV}[1]{\left\langle{#1}\right\rangle}

\def\grtsim{\,\,\rlap{\raise 3pt\hbox{$>$}}{\lower 3pt\hbox{$\sim$}}\,\,}
\def\lsim{\,\,\rlap{\raise 3pt\hbox{$<$}}{\lower 3pt\hbox{$\sim$}}\,\,}

\def\Mp{\bar{M}_{\rm Pl}}
\def\beq{\begin{equation}}
\def\eeq#1{\label{#1}\end{equation}}
\def\eeqn{\end{equation}}
\def\beqa{\begin{eqnarray}}
\def\eeqa#1{\label{#1}\end{eqnarray}}
\def\eeqan{\end{eqnarray}}
\def\CR{\nonumber \\ }
\def\leqn#1{(\ref{#1})}

\def\n{{(n)}}
\def\tofro{\leftrightarrow}
\def\dm{\delta m^2}

\begin{document}
\begin{flushright}
LBNL-49369\\
UPR-0977T\\
January 2002
\end{flushright}
\bigskip\bigskip

\thispagestyle{empty}
\flushbottom

\centerline{\Large\bf Constraints on Large Extra Dimensions from} \vspace{10pt} \centerline{\Large\bf Neutrino
Oscillation Experiments} \vspace{22pt}

\centerline{\bf H. Davoudiasl$^{1, a}$, P. Langacker$^{1, b}$\footnote{Permanent address: Department of Physics
and Astronomy, University  of Pennsylvania,  Philadelphia, PA 19104.}, and M. Perelstein$^2$} \vspace{8pt}
\centerline{\it $^1$   School of Natural Sciences, Institute for Advanced Study, Princeton, NJ 08540}
\centerline{e-mail: $^a$ hooman@ias.edu, $^b$ pgl@electroweak.hep.upenn.edu}
\centerline{\it $^2$  Theoretical   Physics  Group, Lawrence Berkeley National Laboratory, Berkeley, CA 94720}

\centerline{e-mail: meperelstein@lbl.gov}

\vspace*{0.9cm}

\begin{abstract}

The existence of bulk sterile neutrinos in theories with large
extra dimensions can naturally explain small 4-dimensional Dirac
masses for the  active neutrinos.  We study a model with 3 bulk
neutrinos and derive various constraints on the size of the
extra  dimensions from neutrino oscillation experiments.  Our
analysis includes recent solar and atmospheric data from SNO and
Super-Kamiokande, respectively, as well as various reactor and
accelerator experiments.  In addition, we comment on possible
extensions of the model that could accommodate the LSND results,
using natural parameters only.  Our most conservative bound,
obtained from the atmospheric data in the hierarchical mass
scheme, constrains the largest extra dimension to have a radius
$R < 0.82 \, \mu$m. Thus, in the context of the model studied
here, future gravitational experiments are unlikely to observe
the effects of extra dimensions.

\end{abstract}

\newpage

\section{Introduction}

The substantial body of theoretical work done on models with
Large Extra Dimensions (LED's) has provided us with new insights
on some old problems. Even though the original motivations were
mostly related to the question of gauge hierarchy~\cite{ADDA},
various other applications have been found and studied. Among
these is an explanation for small but non-vanishing neutrino
masses~\cite{DDG}. In the more traditional scenario with a large
energy desert between the scales of electroweak symmetry breaking
and grand unification, the small Majorana masses can be easily
generated through the see-saw mechanism. In the models with
LED's, however, this is not possible, since physics at energy
scales well above a TeV is no longer described by a renormalizable
quantum field theory. There is, however, an
alternative mechanism. In these models, all the Standard Model
(SM) particles, including left-handed neutrinos, have to be
confined to a 4-dimensional subspace (3-brane) inside the full
space-time. On the other hand, if an SM singlet fermion (such as
the right-handed neutrino) is present, it can propagate in more
than four dimensions. The large volume of the extra dimensions
leads to a suppression of the wave function of this fermion on the
brane, which in turn allows small Dirac neutrino masses to be
generated naturally. This point contributes positively towards a
comparison of the models with LED's with those of the traditional
energy desert paradigm.

From the four-dimensional point of view, a higher-dimensional SM
singlet fermion can be decomposed into a tower of Kaluza-Klein
(KK) excitations. These states do not have SM charges, and can
therefore be classified as ``sterile''. However, the KK states do
not completely decouple from the system: there are mixings
between them and the lowest-lying, active neutrinos. Thus, they
can participate in neutrino oscillations, acting effectively as a
large number of sterile neutrinos. The implications of this
picture have been studied in a number of
papers~\cite{DDG,DS,MPL,BCS,Apostolos}\footnote{There have been a number of
other models which combined the ideas of LED with additional
ingredients, such as small Majorana masses for the brane
neutrinos~\cite{othermodels}.}.

The recent experimental results of the SNO collaboration~\cite{SNO}, 
in conjunction with the data from Super-Kamiokande~\cite{SKsolar}, 
have yielded 
strong constraints on the contribution of any sterile state to
the solar neutrino anomaly. Motivated by the new data, we have
reexamined the phenomenological constraints on models with
extra-dimensional neutrinos. We have considered the bounds coming
from reactor and accelerator experiments, as well as measurements
of fluxes of solar and atmospheric neutrinos. We have studied a
model~\cite{MPL,BCS} with 3 bulk neutrinos which have Yukawa
couplings to the 3 active brane neutrinos. We will refer to this
model as the (3,~3) model, where the first integer corresponds to
the number of brane neutrinos and the second is the number of
bulk neutrinos. Unlike most previous studies, we are considering
the situation in which the dominant effects for the solar and
atmospheric neutrinos are oscillations amongst the three active
states, with oscillations into the sterile KK excitations
regarded as a perturbation to be constrained by the data. This
framework, which is motivated by the SNO and Super-Kamiokande
data, allows us to obtain simple analytic results throughout the
analysis. We emphasize that the principal role of the extra
dimensions in this case is to provide a natural framework for
small Dirac neutrino masses.

In this paper, we will assume that the manifold on which the
extra dimensions are compactified is highly asymmetric, with one
dimension much larger than the rest. Unlike most collider,
astrophysical and cosmological bounds, which are largely
independent of the compactification manifold's shape and only
constrain its volume, the analysis of this paper will yield
bounds on the size of the largest dimension. The bounds we obtain
are sufficiently tight to rule out (in the context of the model
considered here) effectively the possibility that the near-future
Cavendish type experiments will observe a signal for large extra
dimensions. Although this conclusion clearly depends on the assumed 
properties of the neutrino sector of the theory, we note that the
(3,~3) model seems to be the simplest and most natural way to 
incorporate small neutrino masses in theories with LED's. 

Along with the solar and atmospheric neutrino anomalies, evidence
for neutrino oscillations has been reported by the LSND
experiment~\cite{LSND}. The LSND results cannot be accommodated
within the minimal (3,~3) model without the introduction of {\it
ad hoc} parameters, such as brane Majorana masses. We propose two
extensions of the (3,~3) scheme which can address these results,
using only the natural parameters of the model.  The first
proposal introduces two new sterile states, one on the brane and
the other in the bulk. Thus, this model is of the (4,~4) type.
The second proposal uses an additional LED of a different radius,
in which a sterile bulk state is sequestered.

The paper is organized as follows. In the next section, we
describe our setup and introduce the necessary notation.  The
motivation behind our analysis as well as our approach and choice
of parameters are discussed in section \ref{sec:SNO}.  Section
\ref{sec:exp} discusses constraints on the model from reactor,
accelerator, atmospheric, and solar data. We present our results
in the context of the hierarchical, inverted, and degenerate mass
schemes. In section \ref{sec:LSND}, we address the LSND results
and outline the aforementioned proposals for accommodating them.
Our conclusions are presented in section \ref{sec:conc}.

\section{Formalism and Notation}

\subsection{General Setup}

As required in theories with LED's, we will assume that the
Standard Model (SM) fields, including the three families of
left-handed neutrinos $\nu_L^\alpha$ ($\alpha=e, \mu, \tau$) and
the Higgs doublet $H$, are confined to a four-dimensional brane,
while gravitational fields propagate in a space-time with
$\delta\ge2$ additional compactified dimensions of volume
$V_\delta$. The fundamental gravitational scale of the
higher-dimensional theory $M_F$ is generally taken to be close to
the weak scale to solve the hierarchy problem. To reproduce the
measured strength of four-dimensional gravity at large distances,
we require \beq \bar{M}_{\rm Pl}^2 = M_F^{\delta+2}\, V_{\delta},
\eeq{planck} where $\bar{M}_{\rm Pl} \simeq 2 \times 10^{18}$ GeV
is the reduced Planck mass.

We postulate the existence of three families of SM singlet bulk
fermions $\Psi^\alpha$, which can propagate in 4+$\delta$
dimensions. Yukawa couplings of $\Psi^\alpha$ to the left-handed
neutrinos on the brane give rise to masses for active neutrino
species which are naturally of the right size to accommodate the
solar and atmospheric neutrino anomalies. The same couplings lead
to mixings between the active species and the higher Kaluza-Klein
(KK) components of the bulk fermions, which from the
four-dimensional point of view appear as sterile neutrinos.

Throughout this paper, we will assume that one of the dimensions
is compactified on a circle of radius $R$ which is much larger
than the sizes of the other dimensions. In this case, only the KK
excitations of the bulk neutrinos corresponding to the largest
dimension will be relevant at low energies, and our treatment
will be essentially five-dimensional. Apart from its simplicity,
this model is interesting for the following reason. LED's could be
discovered by Cavendish type experiments which search for
deviations of the gravitational force from Newton's law at short
distances. At present, these experiments probe distances of order
0.2 mm~\cite{Hoyle}, and sensitivities of order 0.05 mm can be
achieved in the near future~\cite{Long}. If the compactification
manifold is symmetric, astrophysical constraints on the radii of
the extra dimensions imply that these sensitivities will not be
sufficient to detect a signal~\cite{SN1987a}. However, the
astrophysical and other high-energy constraints primarily
restrict the {\it volume} of the extra dimensions, while the
Cavendish type experiments are sensitive to the size of the {\it
largest} dimension. Thus, in models with highly asymmetric
compactifications, such as the one we are considering, these
experiments still have a chance of observing deviations from
Newton's law. Unfortunately, the constraints obtained in this
paper indicate that even this possibility may be already ruled
out, if the neutrino sector of the model has the properties
assumed in our study.

As we will see below, the simple setup described here cannot accommodate 
the positive result reported by the LSND neutrino oscillation experiment. 
In section 5, we will propose two simple extensions of this setup that 
might be capable of explaining the LSND result.

\subsection{The (3,~3) Model}
\label{sec:form}

From the 4-dimensional point of view, the 5-dimensional fermion
$\Psi$ can be decomposed into two Weyl fermions, $\psi_L$ and
$\psi_R$. The action of the model is given by \beq S = \int d^4 x
d y \,\,i \bar{\Psi}^\alpha \Gamma_A \partial^A \Psi^\alpha \,\,
+\,\, \int d^4x \left(i\,\bar{\nu}_L^\alpha \gamma_\mu
\partial^\mu \nu_L^\alpha \,+\, \lambda_{\alpha\beta}\,H\, \bar{\nu}_L^\alpha
\psi_R^\beta(x, 0) \,+\,\,{\rm h.c.} \right), \eeq{action} where
$\Gamma_A, A = 0, \ldots, 4$ are the 5-dimensional Dirac matrices.  
Note that this action
conserves lepton number; in particular, we do not introduce
Majorana masses for the left-handed neutrinos. The Yukawa
couplings $\lambda_{\alpha\beta}$ have dimensions of
(mass)$^{-\delta/2}$. Since the only mass scale in the problem at
this point is $M_F$, we introduce dimensionless Yukawa couplings
via \beq h_{\alpha\beta} = \lambda_{\alpha\beta} \,
M_F^{\delta/2}. \eeq{yukawa} We will assume that
$h_{\alpha\beta}$ are of order one.

Let us decompose the 5-dimensional fermions $\psi_{L, R}$ into a
tower of KK states $\psi_{L,R}^\n$, $n=-\infty\ldots\infty.$  It
turns out that certain linear combinations of the KK states are
completely decoupled from the left-handed neutrinos, and
therefore need not be considered. The states that do couple are
given by \beqa \nu_R^{\alpha(0)} &\equiv& \psi_R^{\alpha(0)}; \CR
\nu_R^{\alpha(n)} &=& \frac{1}{\sqrt{2}}\,(\psi_R^{\alpha(n)}+
\psi_R^{\alpha(-n)}), \,\,n=1\ldots \infty; \CR \nu_L^{\alpha(n)}
&=& \frac{1}{\sqrt{2}}\,(\psi_L^{\alpha(n)}+
\psi_L^{\alpha(-n)}), \,\,n=1\ldots \infty. \eeqa{states} In this
notation, the mass terms resulting from \leqn{action} take the
form \beq m^D_{\alpha\beta} \left(\bar{\nu}_R^{\alpha(0)}
\nu_L^\beta \,+\, \sqrt{2} \, \sum_{n=1}^\infty
\bar{\nu}_R^{\alpha(n)} \nu_L^\beta \right) \,+\,\sum_{n=1}^\infty
\frac{n}{R}\,\bar{\nu}_R^{\alpha(n)} \nu_L^{\alpha(n)}\,+\,\,{\rm
h.c.} \eeq{massterms} The Dirac mass matrix is given by \beq
m^D_{\alpha\beta} = h_{\alpha\beta} (v M_F/\Mp), \eeq{massmatrix}
where $v \equiv \VEV H = 246$ GeV, and we have used
Eq.~\leqn{yukawa}, $(M_F^\delta V_\delta)^{1/2} = \bar{M}_{\rm
Pl}/M_F$ (note that the KK modes have a prefactor proportional to
$V_\delta^{-1/2}$), and Eqs. \leqn{states}. If $M_F/\Mp \ll 1$,
as is natural in models with LED's, the common scale of Dirac
masses is well below the electroweak symmetry breaking scale $v$
even for order-one Yukawa couplings. For example, with $M_F \sim
100$ TeV, and $0.1 \lsim h_{\alpha\beta} \lsim 1$, Dirac masses
of the active species will coincide with those required by the
solar and atmospheric data. This is the higher-dimensional
version of the see-saw mechanism.

Diagonalization of the mass terms in \leqn{massterms} can be performed in two steps. First, we find $3\times3$
matrices $l$ and $r$ which diagonalize the Dirac mass matrix, $m_D^d \equiv r^\dagger m^D l = {\rm\,diag\,}(m^D_1,
m^D_2, m^D_3).$ We define \beqa \nu_L^\alpha &=& l^{\alpha i} \nu_L^{\prime\,i};   \CR \nu_R^{\alpha(n)} &=&
r^{\alpha i} \nu_R^{\prime\,i(n)}, \,\, n=0 \ldots\infty; \CR \nu_L^{\alpha(n)} &=& r^{\alpha i}
\nu_L^{\prime\,i(n)}, \,\, n=1 \ldots\infty. \eeqa{newstates1}
The transformation of the $\nu_L^{\alpha(n)}$ by $r^{\alpha i}$ greatly simplifies the problem by leaving the
second term in \leqn{massterms} diagonal.
Then, \leqn{massterms} takes the form \beq
\sum_{i=1}^3 {\bar{\nu}}^{\prime\,i}_R \,M_i\, \nu^{\prime\,i}_L \,+\, {\rm h.c.} \eeq{massterms1} where
$\nu^{\prime\,i}_L = (\nu^{\prime\,i}_L, \nu^{\prime\,i(1)}_L, \nu^{\prime\,i(2)}_L, \ldots)^T$,
$\nu^{\prime\,i}_R = (\nu^{\prime\,i(0)}_R, \nu^{\prime\,i(1)}_R, \nu^{\prime\,i(2)}_R, \ldots)^T$, and $M_i$ is
an infinite-dimensional matrix given by \beq M_i = \left(\begin{array}{cccc} m^D_i & 0 & 0 & \ldots \cr
                         \sqrt{2}m^D_i & 1/R & 0 & \ldots \cr
                         \sqrt{2}m^D_i & 0 & 2/R & \ldots \cr
                         \ldots & \ldots & \ldots & \ldots
      \end{array}\right).
\eeq{mmatrix}

To complete the diagonalization, we need to find
infinite-dimensional matrices $L$ and $R$ such that $R_i^\dagger
M_i L_i$ is diagonal; then, the mass eigenstates are given by
$\tilde{\nu}^i_L = L_i^\dagger\,\nu^{\prime\,i}_L$,
$\tilde{\nu}^i_R = R_i^\dagger\,\nu^{\prime\,i}_R.$ In this
paper, we are primarily interested in the decomposition of the
flavor-basis brane neutrino states $\nu_L^\alpha$ in terms of the
mass eigenstates. These are given by \beq \nu_L^\alpha =
\sum_{i=1}^3 l^{\alpha i} \sum_{n=0}^\infty L_i^{0n}
\tilde{\nu}^{i(n)}_L. \eeq{decomp} The easiest way to find $L$ is
to observe that it has to diagonalize the Hermitian matrix
$M^\dagger M$, and therefore consists of its eigenvectors. This
procedure was performed in Refs.~\cite{DDG, DS, BCS}. The
component of  $L$ that enters Eq.~\leqn{decomp} is given by \beq
(L_i^{0n})^2 = {2\over{1 + \pi^2\xi_i^2/2 + 2
\lambda_i^{(n)2}/\xi_i^2}}, \eeq{lmatrix} where $\xi_i= \sqrt{2}
m^D_i R$, and $\lambda_i^{(n)2}$ are the eigenvalues of the matrix
$R^2 M^\dagger M$, which are found from the equation \beq
\lambda_i \,-\, \frac{\pi}{2}\,\xi_i^2\,\cot(\pi \lambda_i) = 0.
\eeq{eigen} The mass of the mode $\tilde{\nu}^{i(n)}_L$ is simply
given by $\lambda_i^\n/R$. Similarly, \beq L_i^{kn} = \frac{k
\xi_i}{\lambda_i^{(n)2} - k^2}  L_i^{0n}, \eeq{offdiag} where
$k=1 \ldots \infty$ and $n= 0  \ldots \infty$.

Since our philosophy will be to consider the effects of extra
dimensions as small perturbations on top of the oscillations
amongst zero-modes, we will be mostly interested in the case where
$R^{-1} \gg m^D_i$, that is, the limit of small $\xi_i$. In this
limit, we find \beqa \lambda_i^{(0)} &=&
\frac{\xi_i}{\sqrt{2}}\,\left(1-\frac{\pi^2}{12}\xi_i^2
+\ldots\right), \hskip1cm L_i^{00} = 1 - \frac{\pi^2}{12}\xi_i^2
+ \ldots \CR \lambda_i^{(k)} &=& k + {1 \over 2k} \xi_i^2 +
\ldots, \hskip2cm L_i^{0k} = {\xi_i \over k} + \ldots \CR
L_i^{k0} &=& -{\xi_i \over k} + \ldots, \hskip2cm L_i^{kk} = 1 -
\frac{\xi_i^2}{2 k^2} + \ldots,
 \eeqa{smallxi}
and $L_i^{jk} = O(\xi_i^2)$,  where  $j \ne k=1 \ldots \infty$.
A quantity of crucial interest is the probability of finding a
neutrino of flavor $\beta$ in a beam that was born with flavor $\alpha$ and has travelled a distance $L$. This
probability is given by \beqa P_{\alpha\beta}(L) &=& |A_{\alpha\beta}(L)|^2, \CR A_{\alpha\beta}(L) &=&
\sum_{i=1}^3 l^{\alpha i} l^{\beta i *} A_i(L), \eeqa{probab} where $E_\nu$ is the energy of the beam, and \beq
A_i(L) = \sum_{n=0}^\infty (L_i^{0n})^2 \exp\left(i {\lambda_i^{(n)2} L \over 2 E_\nu R^2}\right). \eeq{ai}
Similarly, the probability for $\nu_\alpha$ to oscillate into sterile neutrinos is
\beq
P_{\alpha s}(L) = \sum_{i=1}^3 \sum_{k=1}^\infty |B_{\alpha i(k)} |^2,
\eeq{psterile}
where \beq B_{\alpha i(k)} = l^{\alpha i} \sum_{n=0}^\infty
L_i^{0n} L_i^{kn} \exp\left(i {\lambda_i^{(n)2} L \over 2 E_\nu R^2}\right).
\eeq{bdef}
Of course, $\sum_{\beta=1}^3  P_{\alpha\beta}(L) + P_{\alpha s}(L) =1$.

\section{LED Sterile Neutrinos and SNO: Motivation and Choice of Parameters}
\label{sec:SNO}

The flux of solar neutrinos has been measured by several
collaborations using independent experimental
techniques~\cite{review}. The observed flux is consistently lower
than the predictions of the Standard Solar Model (SSM) by a factor of two
to three, depending on the spectral sensitivity of the experiment. The 
most elegant explanation of this deficit is
provided by the hypothesis of neutrino oscillations. The
neutrinos produced in the Sun are in the flavor ($\nu_e$)
eigenstate. If this state is a nontrivial superposition of the
mass eigenstates, some of the neutrinos would change flavor on
the way to the Earth. Because the terrestrial experiments are
predominantly sensitive to electron neutrinos, this effect leads
to a suppression of the measured fluxes.

Until recently, the experimental data could be explained either
by an oscillation of $\nu_e$ into other active species, $\nu_\mu$
or $\nu_\tau$, or by its oscillation into a sterile neutrino
sector, denoted by $\nu_s$. However, the best fit was for
oscillations into active neutrinos, especially when the
Super-Kamiokande day-night and spectrum information was
included~\cite{SKsolar}. The case has recently been greatly
strengthened. The SNO collaboration~\cite{SNO} has unambiguously
demonstrated the presence of non-electron active neutrinos in the
solar flux. Moreover, the total flux of solar neutrinos implied
by SNO and Super-Kamiokande data is in excellent agreement with
the SSM calculations~\cite{SSM}.  These results mean that only a
 fraction (if any) of the solar neutrinos get converted into sterile species.

In our framework, there are three active species, namely the left-handed brane neutrinos $\nu_L^\alpha$. These
states can mix both among themselves and with the excited KK states of the bulk neutrinos, $\nu_L^{\alpha(n)}$.
The bulk neutrinos do not carry any SM charges, and therefore are sterile. Thus, the SNO data indicates that the
solar neutrino deficit is primarily due to the oscillations amongst the active brane states,
and allows one to put
constraints on their mixings with the sterile KK states.

In the model proposed by Dvali and Smirnov~\cite{DS}, the solar neutrino deficit is explained by the oscillations
of $\nu_L^e$ into the KK excitations of the bulk neutrino of the same flavor. The other active flavors,
$\nu_L^\mu$ and $\nu_L^\tau$, play no role. (In the language of section~\ref{sec:form}, this
corresponds to $l^{e1}=1, l^{e2}=l^{e3}=0.$) It is immediately obvious that this model is in contradiction with the
SNO data: it predicts that no non-electron active neutrinos should be observed in the solar flux. To accommodate
the data, it is necessary to introduce non-zero mixing among the active neutrinos.

Motivated by the SNO data, we will take the following approach.
We assume that in the limit in which the sterile KK neutrinos are
decoupled ($R\to0$ limit) the oscillations amongst the remaining
active species provide a good fit to the solar neutrino data. To
be specific, we assume that the large mixing angle MSW solution to
the solar neutrino problem, which provides the best fit to the
data~\cite{review}, is correct. We fix $\dm_{sol} =
\dm_{21}\equiv (m_2^D)^2-(m_1^D)^2=3.7\times 10^{-5}$ eV$^2$ and
take the mixing matrix elements to be $l^{e1} = \cos\beta,
l^{e2}=\sin\beta$,
 with $\tan^2\beta=0.37$. (Varying these parameters over their allowed ranges does not
significantly affect our conclusions.) Furthermore, we will
assume that the atmospheric neutrino anomaly observed by
Super-Kamiokande~\cite{superK} is also due predominantly to the
oscillations between the active species ($\nu_\mu\tofro\nu_\tau$).
This assumption is supported by the analysis of the
neutral-current-enhanced events, the zenith angle distribution,
and $\tau$-appearance candidates
 in Super-Kamiokande~\cite{superK1}, and by the zenith angle distribution 
observed by MACRO~\cite{MACRO},
which strongly disfavor the possibility that the anomaly is due
to $\nu_\mu\tofro \nu_s$ oscillations. A good fit to the Super-Kamiokande 
data is
obtained if we choose the mixing between $\nu_\mu$ and $\nu_\tau$
to be maximal, and the mass splitting $|\dm_{atm}| =
|\dm_{32}|=3\times 10^{-3}$ eV$^2$. Finally, we (generally)
choose $l^{e3}$ to be 0 because of the strong constraint on this
matrix element provided by a combination of atmospheric neutrino
data and reactor experiments. With these choices, the only
remaining free parameters are the radius of the extra dimension
$R$ and the absolute scale of the Dirac masses. For the latter,
we will consider three possibilities: (1) the``hierarchical''
mass scheme, $m_1^D\approx 0$, $m_3^D\gg m_2^D$, implying $m_3^D
\approx \left( \dm_{atm} \right)^\frac{1}{2} \approx 0.055$ eV and
$m_2^D \approx \left( \dm_{sol} \right)^\frac{1}{2} \approx
0.006$ eV;
 (2) the ``inverted'' mass scheme,
$m_3^D\approx 0$, $m_1^D \approx m_2^D
\approx \left( \dm_{atm} \right)^\frac{1}{2} \approx 0.055$ eV;
 and (3) the ``degenerate'' scheme\footnote{The original motivation for the degenerate scheme, i.e., mixed
hot-cold dark matter models, has now largely disappeared, but it is still a logical possibility.
Note also that this scheme requires a fundamental scale $M_F$ larger than 100 TeV.},
with $m_1^D \approx m_2^D \approx m_3^D
\approx 1$ eV, where the common value has
been chosen to be close to the cosmological limit of $\sim 4.4$ eV
on the sum of the light neutrino masses~\cite{MDM}. In the next section, we will present experimental
constraints on the radius
$R$ in each of these three schemes.

An important advantage of our perturbative approach is that it
allows us to obtain simple analytic results throughout the
analysis, making the underlying physical picture quite
transparent. A more complete approach would be to fit the data
allowing all the parameters of the model (3 Dirac masses, 3
mixing angles, 1 CKM phase and the radius of the extra dimension
$R$) to vary, and then obtain bounds on $R$ by marginalizing over
the other variables. The resulting bounds would necessarily be
less restrictive than the results of our present approach.
However, as we explained above, current experimental data
disfavors the possibility that oscillations into sterile states
play a dominant role for either solar or atmospheric neutrinos.
This implies that the results presented here should provide a
good first approximation which could then be refined by a more
systematic analysis.

\section{Experimental Constraints on $R$ from Neutrino Oscillation Data}
\label{sec:exp}

Having stated our approach, we will use the formalism developed in the previous two sections to derive constraints
on the radius of the extra dimension $R$ from a variety of sources, such as reactor and accelerator searches for
neutrino oscillations and solar and atmospheric neutrino flux measurements.  We will next consider each of these
constraints in turn.

\subsection{Reactor and Accelerator Experiments}

Large fluxes of anti-electron neutrinos are produced at nuclear
power  reactors. If the flux  can be either predicted accurately
or measured by a nearby detector, measuring the $\nu_e$ flux at a
certain distance $L$ from the reactor gives the electron neutrino
survival probability $P_{ee}(L)$. Several experiments, such as
CHOOZ~\cite{CHOOZ} and Bugey~\cite{Bugey}, have utilized this
approach to search for neutrino oscillations. Their results are
consistent with no oscillation hypothesis, that is, $P_{ee}(L)
\simeq 1$.

The electron neutrino survival probability is determined by
Eq.~\leqn{probab}: \beq P_{ee}(L) = \left| \sum_{i=1}^3 |l^{e
i}|^2 A_i(L) \right|^2, \eeq{survive} where  $A_i$ is given by
Eq.~\leqn{ai}. For $l^{e3} \simeq 0$ and the values chosen for the
$m^D_i$, $P_{ee}(L) \approx 1 - P_{es}(L)$. For the case of small
$\xi$ that we will be primarily interested in, \beq P_{es}(L)
\approx \sum_i 2 |l^{ei}|^2 \xi_i^2 \sum_{k=1}^\infty \left(
\frac{1 - \cos \phi^k_i}{k^2} \right) + O(\xi^4), \eeq{oscprob}
where the oscillation phase is \beq \phi^k_i \equiv \frac {
(\lambda_i^{(k)2}  - \lambda_i^{(0)2}) L}{ 2 E_\nu R^2} \approx
\left( \frac{k^2}{R^2} - (m_i^{D})^2 \right) \frac {L}{ 2
E_\nu}.\eeq{oscphase} It is convenient to evaluate the
probabilities numerically, keeping only a finite number of terms
in the sums in Eq.~\leqn{ai} or Eq.~\leqn{oscprob}. (The
oscillation probability into the $k$-th KK state is suppressed by
$k^{-2}$ for large $k$.)

For the CHOOZ experiment, the distance from the reactor to the detector is $L\approx 1$ km, and the  neutrino
energies $E_\nu$ range from about $(1-6)$ MeV. The ratio of the measured flux to the no-oscillation Monte Carlo
prediction reported by CHOOZ, ${\cal R} =1.01\pm0.028$(stat.)$\pm0.027$(syst.),\ implies\footnote{These results and
similar ones throughout the paper are obtained by
the conservative method of assuming a Gaussian distribution for $P_{ee}(L)$ in the physical
region $0 \le P_{ee}(L) \le 1$, using the central value and uncertainty from ${\cal R}$ but renormalizing so that
the total probability for finding any value of $P_{ee}(L)$ in the physical region is unity.} \beq P_{ee}(L) >
0.942 \ \ \ \ [P_{es} < 0.058] \eeq{choozbnd} at 90\% c.l. Taking an average value $L/E_\nu \sim 300$ m/MeV, we
obtain the following 90\% c.l. constraints on the radius of the extra dimension: (1) in the hierarchical scheme,
$\xi_2<0.43$, corresponding to $1/R > 0.02$ eV; (2) in the inverted scheme, $\xi_1 \approx \xi_2 < 0.13$, $1/R >
0.60$ eV; and (3) in the degenerate scheme, $\xi_1\approx\xi_2\approx\xi_3<0.13$, $1/R > 10.9$ eV. We present all
the corresponding upper bounds on $R$ in Table~\ref{tab:rabounds}.  The constraints in the inverted and degenerate
schemes are quite a bit stronger than in the, perhaps more natural, hierarchical scheme. There are two reasons for
this. Firstly, since $\xi=\sqrt{2}mR$, the same bound on $\xi$ results in stronger bounds on $R$ if the active
neutrinos are heavy. Secondly, with the parameters of the CHOOZ experiment, the phases of the oscillations into
the first few KK states are small in the hierarchical case, suppressing the oscillations. The higher modes have
larger oscillation phases but small mixings.

These results were obtained numerically using
\leqn{ai}, but it is instructive to discuss the approximate formula \leqn{oscprob}, which yields almost
identical results, even in the hierarchical mass scheme.
In the hierarchical case,
\beq
P_{es}(L) \approx 2 \sin^2 \beta \ \xi_2^2 \sum_{k=1}^\infty \left( \frac{1 - \cos \phi^k_2}{k^2} \right).
\eeqn
The phases $\phi^k_2$ of the first few KK states are not large, so  the
$\cos \phi^k_2$ terms must be kept. We have verified that integrating over the neutrino energy $E_\nu$, which
smooths the oscillations, gives results almost identical to using the average $L/E_\nu$.
In the inverted and degenerate cases, the $\phi^k_{1,2}$ are large and the
$\cos \phi^k_{1,2}$
average to zero. Since $\xi_1 \approx \xi_2$ in these cases, one finds $P_{es}(L) \approx
\xi_2^2 \pi^2/3$.

Our assumption $l^{e3} = 0$ yields the most conservative limit. Relaxing this assumption
would result in
more stringent limits because $1 - P_{ee}(L)$ would receive new
positive contributions $\sim |l^{e3}|^2 \left( 2 + \frac{\pi^2}{3} \xi_3^2 \right)$, where the two terms
are respectively due to oscillations into active neutrinos and into the third KK tower.

The analysis of the Bugey experiment is very similar. The results are presented in Table~\ref{tab:rabounds}.
Because of the small distance between the reactor and the detector in this experiment, it is not possible to put
significant bounds on $R$ in the hierarchic  scheme: oscillation phases are substantial only for very
high KK modes, but the mixings of those modes with the active neutrino are severely suppressed. In the other two
mass schemes, the bounds are similar to CHOOZ, although somewhat weaker due to a lower precision of the flux
measurement.

Neutrino oscillations can also be searched for at accelerators, where beams of muon neutrinos can be produced. The
CDHS experiment~\cite{CDHS} at CERN has searched for the disappearance of muon neutrinos
from a beam with
average energy $E_\nu\approx 3$ GeV, by measuring the
ratios of the numbers of events in two detectors  at 130 and 885 m from the source. The negative results of this
experiment can, in principle, yield bounds on
$R$.  However, for the $L/E_\nu$ range of the experiment we obtain significant
bounds only for the degenerate scheme. Averaging over the neutrino energies in the CDHS range
and comparing with the measured ratio,
we obtain the constraint that $1/R $ must be greater than around 6 eV, with a small
allowed window around 4 eV. Because of the imprecise nature of the degenerate solution, we just
quote $1/R >$ 4 eV.

More recently, the CHORUS~\cite{CHORUS} and NOMAD~\cite{NOMAD}
collaborations have searched for an {\it appearance} of tau
neutrinos in a beam originally composed of $\nu_\mu$. The
non-observation of tau neutrino events has led to strong bounds
on the probability of $\nu_\mu\to\nu_\tau$ oscillations, of order
$P<10^{-4}$. Unfortunately, this constraint does not put
significant bounds on $R$. With the parameters of these
experiments, the phases of the oscillations into the first 10-100
modes are small in hierarchical and inverted schemes. For higher
modes, the oscillation amplitudes are of order $\xi^2/n^2$, since
the KK mode has then to oscillate back into $\nu_\tau$ to be
detected.  Thus, the corresponding probability is severely
suppressed: $P_{\mu\tau}\sim \xi^4/n^4$. The resulting
constraints are of the order $\xi\sim 1$ and are not competitive
with bounds from neutrino disappearance experiments. For the
degenerate scheme the oscillation phases could be substantial
even for low-lying modes, but because of unitarity cancellations
the resulting oscillation probabilities are severely suppressed,
and no useful bound can be obtained. (The cancellations appearing
in this case are analogous to those discussed in
section~\ref{sec:LSND}.)

Another accelerator experiment, LSND, has reported evidence for
neutrino oscillations. We postpone the discussion of this result
until section~\ref{sec:LSND}.

\subsection{Atmospheric Neutrinos.}

The Super-Kamiokande experiment~\cite{superK} has measured the
fluxes of electron and muon neutrinos produced in the atmosphere
by cosmic rays. While the flux of electron neutrinos is well
described by a Monte Carlo (MC) simulation based on the
no-oscillation hypothesis, the muon neutrino flux deviates
significantly from this prediction. In particular, the flux of
up-going muon neutrinos is suppressed in comparison with the
down-going flux. The data, including the corresponding zenith
angle distribution, is consistent with the hypothesis that on
their way through the Earth the up-going muon neutrinos undergo
oscillations that convert them into $\nu_\tau$'s. (An alternative
possibility of $\nu_\mu\to\nu_s$ oscillations is strongly
disfavored~\cite{superK1}.) Fitting the data within the
two-flavor scheme yields the mass difference $|\dm_{32}|\approx
3\times 10^{-3}$ eV$^2$, and the mixing angle
$\theta_{23}\approx\pi/4$ (maximal mixing.)

For the case we are considering, the oscillation of $\nu_{\mu,e}$
into sterile KK states will involve large oscillation phases and
therefore lead to a suppression of the overall $\nu_e$ and/or
$\nu_\mu$ fluxes that is independent of the zenith angle. The
Super-Kamiokande collaboration~\cite{superK} have included the
possibility of such $L/E_\nu$--independent flux changes in their
analysis to take into account the theory uncertainties in the
absolute and relative flux calculations. However, their results
can also be used to constrain $L/E_\nu$--independent oscillations
into sterile states, provided the theoretical uncertainties in
the fluxes are properly included.

The best-fit values of the Super-Kamiokande data~\cite{superK},
assuming a $\nu_\mu \leftrightarrow \nu_\tau$ analysis, suggest
that there is a slight overall excess in the ratio of the
measured to the central value of the simulated number of events.
The data also yield a deficit for the ratio of the $\mu$-like to
$e$-like events, assuming the same two-flavor analysis. In the
analysis of the data~\cite{superK}, the overall excess and the
deficit mentioned above are parameterized by $\alpha = 0.034 \pm
0.25$ and $\beta_s = -0.059 \pm 0.08$, respectively, where the
observed   sub-GeV $\nu_e$ and $\nu_\mu$ rates relative to the
number expected for no oscillations into sterile states are \beqa
r_{\nu_e}  & = & \left( 1 + \alpha \right) \left( 1 -
\frac{\beta_s}{2} \right) \CR r_{\nu_\mu}  & = & \left( 1 +
\alpha \right) \left( 1 + \frac{\beta_s}{2} \right). \eeqan
  The subscript $s$ refers
to the fact that this value is obtained from the sub-GeV events, which yield the most stringent constraints.
  We have assumed
 that the errors are dominated by the theoretical flux
 uncertainties. Quantitatively, the experimental results imply \beq
-\alpha =
\frac{1}{2}[P(\nu_\mu \to \nu_s) + P(\nu_e \to \nu_s)] \eeq{mu+e} and \beq \beta_s = P(\nu_e \to \nu_s) -
P(\nu_\mu \to \nu_s). \eeq{mu-e} In what follows, we use the one-sided 90\% c.l. upper bounds on $-\alpha$ and $\pm
\beta_s$ to derive our constraints.

In the hierarchical mass scheme, the active neutrinos predominantly mix with the $\tilde{\nu}_L^{3(k)}$, since
$\xi_3 \gg \xi_2,\,\xi_1 \approx 0$. Since $l^{e3} = 0$, this mixing can only occur for the muon neutrinos,
decreasing their flux compared to the best-fit value; the electron neutrino flux remains unchanged.
Eq.~\leqn{mu-e} implies that at 90\% c.l.,
\beq P(\nu_\mu \to \nu_s) - P(\nu_e \to \nu_s) <
0.17. \eeq{upper}  Thus, ignoring the $\nu_e$ oscillations, and in the limit $\phi^k_3 \gg 1$, we
obtain $ P(\nu_\mu
\to \nu_s) = \xi_3^2 \pi^2/6 < 0.17$, which yields $1/R > 0.24$ eV.

In the inverted mass scheme with $\xi_1 \approx \xi_2 = \xi$ and $\xi_3 \approx 0$, two KK towers,
$\tilde{\nu}_L^{1(k)}$ and $\tilde{\nu}_L^{2(k)}$, contribute to the oscillations. Both electron and muon
neutrinos can mix with these states, so both fluxes are reduced. It turns out that the suppression is more
significant for the electron neutrino, so we derive our constraint from the upper  bound on
$\beta_s$.  Again,
assuming the phase energy averaging, we find $P(\nu_e \to \nu_s) = 2 P(\nu_\mu \to \nu_s) = \xi^2 \pi^2/3$.  Thus,
 $\xi^2 \pi^2/6 < 0.1$, which yields $1/R > 0.32$ eV.

Finally, in the degenerate mass scheme, with $\xi_1 \approx \xi_2 \approx \xi_3 = \xi$, all three KK
towers can mix with the active neutrinos with an approximately equal strength. In this case, the oscillations into
the KK states suppress the electron and muon neutrino fluxes by the {\it same amount}.
In this case, we use Eq.~\leqn{mu+e} to find $(1/2)[P(\nu_\mu \to \nu_s) + P(\nu_e \to \nu_s)] < 0.39$,
at 90\% c.l.  We have $P(\nu_\mu \to \nu_s) = P(\nu_e \to \nu_s) = \xi^2 \pi^2/3$, where phases are energy
averaged. This yields $\xi^2 \pi^2/3 < 0.39$ and hence $1/R > 4.1$ eV.  As we can see, the bounds from the
atmospheric data are quite stringent.  These results are summarized in Table~\ref{tab:rabounds}.

\subsection{Solar Neutrinos}

As we already discussed in section~\ref{sec:SNO}, the results of
the SNO collaboration~\cite{SNO} make it unlikely that
oscillations into sterile KK neutrinos play a major role in
explaining the solar neutrino deficit. In particular, combining
the SNO~\cite{SNO} and Super-Kamiokande~\cite{SKsolar} results,
and using the new SSM estimate of the flux of $^8B$
neutrinos~\cite{SSM} that incorporates a recent more accurate
measurement of the $^7Be(p,\gamma)^8B$ reaction~\cite{Junghans},
one obtains \beq \sum_{\alpha=\mu,\tau}P(\nu_e\to\nu_\alpha) =
0.62 \pm 0.21, \ \ \ P(\nu_e\to\nu_s) = 0.083 \pm 0.21, \ \ \
P(\nu_e\to\nu_e) = 0.30 \pm 0.05 \eeq{solarfluxes} for the
probabilities for an initial  $^8B$ neutrino in the
SNO/Super-Kamiokande energy range to oscillate into $\nu_\mu$,
$\nu_\tau$, or a sterile state, or to remain a $\nu_e$,
respectively. This leads to \beq P(\nu_e\to\nu_s) < 0.40
\eeq{solarlimit} at 90\% c.l. (The corresponding limit would be
0.41 using the older cross sections.) \if In fact, our entire
approach of fixing the parameters of the oscillations between the
active species was motivated by this observation. In this
subsection, we discuss quantitatively the SNO constraint on the
radius of the extra dimension $R$ within our framework. \fi

To compute the fluxes of solar neutrinos, we need to take into
account matter effects in the Sun. In vacuum, the flavor-basis
active neutrino states can be decomposed into the mass
eigenstates according to \beq \nu_\alpha =
\sum_{i=1}^3\sum_{n=0}^\infty U_{\alpha i}^{(n)}
\tilde{\nu}^{i(n)}\,. \eeq{Udef} (This is just Eq.~\leqn{decomp}
of section~\ref{sec:form}; we have dropped the subscript $L$ to
avoid cluttering and defined $U_{\alpha i}^{(n)} = l^{\alpha i}
L_i^{0n}.$) In matter, an additional, flavor-dependent effective
mass term is generated for active neutrino species. The
decomposition \leqn{Udef} is modified by this effect: \beq
\nu_\alpha = \sum_{i=1}^3\sum_{n=0}^\infty U_{\alpha i}^{{\rm
m}\,(n)} \tilde{\nu}_{\rm m}^{i(n)}, \eeq{Udefm} where
$\tilde{\nu}_{\rm m}^{i(n)}$ and $ U_{\alpha i}^{{\rm m}\,(n)}$
are respectively the eigenstates and mixing matrix elements in
matter. These of course depend on the  matter density, and
therefore change as the neutrino travels through the Sun. We
assume that neutrino propagation is adiabatic, that is, the
fraction of each of the mass eigenstates in the beam remains
constant. This is an excellent approximation for the large mixing
angle (LMA) solar neutrino parameters that we are
using~\cite{review,KP}. Since all the phases associated with
solar neutrino oscillations are large, we can neglect the
interference effects between the matter eigenstates. Then, the
oscillation probability is given by a simple formula~\cite{KP}
\beq P(\nu_e\to\nu_f) = \sum_{i,n} P(\nu_e\to\tilde{\nu}_{\rm
m}^{i(n)}) \cdot P(\tilde{\nu}^{i(n)}\to\nu_f) \,=\, \sum_{i,n}
|U_{e i}^{{\rm m}\,(n)}|^2\,|U_{f i}^{(n)}|^2, \eeq{probsol} In
this equation and below, $\tilde{\nu}_{\rm m}^{i(n)}$ and $
U_{\alpha i}^{{\rm m}\,(n)}$ are evaluated at the neutrino
production point in the core of the Sun, and we have assumed that
the neutrino flux is measured in the vacuum (as is the case for
solar neutrinos.)

The values of the mixing angles $U_{ei}^{{\rm m}\, (n)}$ in \leqn{probsol} can be computed exactly if
the neutrino energy and electron and neutron densities in the region of the Sun where the neutrinos are created
are known. This in turn depends on
the number and properties of the Mikheyev-Smirnov-Wolfenstein (MSW) resonances~\cite{KP}
passed by the neutrino on its way to the vacuum.

For the energies of the $^8B$ neutrinos observed by
Super-Kamiokande and SNO, an MSW resonance can only occur for
mass-squared splittings less than about $10^{-4}$ eV$^2$. Hence,
in our case, resonant conversion is only possible between the two
zero-mode mass eigenstates, $\tilde{\nu}^{1(0)}$ and
$\tilde{\nu}^{2(0)}$. The higher mass states, which are mainly
the sterile KK excitations, do not undergo resonant conversion
and are little affected by matter effects for the masses required
by the reactor and atmospheric data. Thus, there is only one
resonance to consider. We will assume that {all} solar neutrinos
observed by SNO and Super-Kamiokande cross this resonance, and
that for all of them the propagation through the resonance is
adiabatic, which are excellent approximations for the LMA
parameter region. The electron neutrino mixing angles in the core of the  
Sun are then 
\beq |U_{e 1}^{{\rm m}\,(0)}|^2 = 0, \hskip1cm |U_{e
2}^{{\rm m}\,(0)}|^2 = |U_{e 1}^{(0)}|^2+|U_{e 2}^{(0)}|^2,
\eeq{inthesun} 
with all the other matrix elements $U_{e i}^{{\rm m}(n)}$ being equal to
their vacuum values.

Formulas \leqn{probsol} and \leqn{inthesun}, together with  \leqn{psterile}
and the small-$\xi$ expansions of the vacuum mixing angles
\leqn{smallxi}, allow us to estimate the sterile component of the solar neutrino flux.
For example, in the hierarchical mass scheme,
\beq P(\nu_e\to\nu_s) = 1 - \sum_{\alpha=e,\mu,\tau} P(\nu_e\to\nu_\alpha) =
\frac{\pi^2}{6}\,(1+\sin^2\beta)\,\xi_2^2. \eeq{sterile}
The experimental bound $ P(\nu_e\to\nu_s) < 0.40$  at 90\% c.l. then implies
 $1/R > 0.02$ eV.
The analysis in the other two mass schemes is analogous except that $ P(\nu_e\to\nu_s) = \pi^2 \xi_2^2/3$, implying
$R > 0.22$ eV (inverted scheme) and $1/R > 4.1$ eV (degenerate scheme).  These results are shown in
Table~\ref{tab:rabounds}.

\begin{table}
\begin{tabular}{cccc}
\hline
\multicolumn{4}{c}{Experimental Bounds} \\
\hline
Experiment  & Hierarchical & Inverted & Degenerate \\
 & (cm, eV)& (cm, eV)& (cm, eV) \\
\hline
CHOOZ & ($9.9 \times 10^{-4}, 0.02$) & ($3.3\times 10^{-5}, 0.60$) & ($1.8\times 10^{-6},10.9 $)\\
BUGEY & none & ( $4.3\times 10^{-5},0.46$)& ($2.4\times 10^{-6},8.3 $) \\
CDHS & none &  none &  ($5 \times 10^{-6},4$) \\
Atmospheric & ($8.2 \times 10^{-5}, 0.24$) & ($6.2\times 10^{-5}, 0.32$) & ($4.8\times 10^{-6}, 4.1$)\\
Solar & ($1.0 \times 10^{-3}, 0.02$) & ($8.9\times 10^{-5}, 0.22$) & ($4.9\times 10^{-6}, 4.1$)\\
\hline
\hline
\end{tabular}
\caption{Upper bounds on $R$ (cm) at 90\% c.l. and the corresponding lower bounds on $1/R$ (eV) from various
measurements.} \label{tab:rabounds}
\end{table}
\vspace{0.2in}

\section{The LSND Experiment}
\label{sec:LSND}

The LSND experiment has searched for ${\bar \nu}_\mu \to {\bar \nu}_e$ oscillations and reports an excess of
${\bar \nu}_e$ events, corresponding to an oscillation probability $P({\bar \nu}_\mu \to {\bar \nu}_e) = (0.264 \pm
0.067 \pm 0.045) \%$~\cite{LSND}.
Similar results are obtained for $\nu_\mu \rightarrow \nu_e$.
 The data suggests that the oscillations occur for 0.2
eV$^2$ $\leq \delta m^2
\leq$ 10 eV$^2$,
with the lower $\delta m^2$ values favored by the nonobservation of an oscillation
signal in the KARMEN experiment~\cite{KARMEN}.
Thus, it seems that mass scales of $\sim 1$ eV must be present in any model that attempts to
describe the LSND results.
In particular, one must introduce a fourth neutrino to account for
LSND as well as the solar and atmospheric results, and this must be a sterile
$\nu_s$ due to the constraint
on the number of light active neutrinos from the $Z$ width~\cite{LEPEWWG}.
Many authors have analyzed such four-neutrino schemes~\cite{sterileschemes}.
The best fits are obtained in the so-called $ 2+2$ schemes, in which there are two closely spaced pairs
or states, with the pairs separated by $\sim 1$ eV and  a small mixing between the pairs to account
for the LSND results. However, the simplest versions, in which  the pairs consist of
either (a) $(\nu_e, \nu_s)$ and $(\nu_\mu, \nu_\tau)$, or
(b) $(\nu_e, \nu_\tau)$ and $(\nu_\mu, \nu_s)$, are now excluded by the solar and
atmospheric data, so that the partners of the $\nu_e$ and $\nu_\mu$ would have
to be mixtures of $\nu_s$ and $\nu_\tau$. The alternative $3+1$ (or $3 +p$, with $p > 1$)
scheme involves three closely-spaced active neutrinos (similar to the 
hierarchical, inverted or degenerate schemes), separated from one or more 
mainly-sterile states by about 1 eV.
$\nu_\mu \rightarrow \nu_e$ oscillations are therefore a sub-leading effect involving the
mixing of both $\nu_\mu$ and $\nu_e$ with the sterile state.
Such schemes are strongly disfavored because it is difficult
to obtain a large enough effect for LSND while still respecting the reactor and accelerator
$\nu_e$ and $\nu_\mu$ disappearance constraints, but are still barely possible.

Although the (3,~3) model contains KK states of mass $\sim 1$ eV in the hierarchical scheme
for large mode
numbers $k \sim 100$, transitions among the active states are suppressed by the smallness
of $|l^{e 3}|$ (so far set to zero) and $\delta m^2$ for the active states, as well as by
the $1/k$ factors in the mixing. Even allowing $|l^{e 3}| \approx 0.2$, calculations based
on our formalism still fall an order of magnitude short of the central value for
$P({\bar \nu}_\mu \to {\bar \nu}_e)$ quoted above. The inverted and the degenerate schemes
result in roughly the same or even smaller probabilities. In generic $3+1$ or $3+p$ schemes
the rate is simply proportional to the fourth power of mixing angles $\theta$ , i.e.,
$P({\bar \nu}_\mu \to {\bar \nu}_e) \propto \theta^2_{e4} \theta^2_{\mu 4}$.
However, in our model there are cancellations between the two KK towers so that
$P({\bar \nu}_\mu \to {\bar \nu}_e) \propto |\xi_2^2 - \xi^2_1|^2 = 4 |\delta m_{sol}^2 R^2|^2$
(or $|l^{e 3}|^2 |\delta m_{atm}^2 R^2|^2$ for  $l^{e 3} \ne 0$). Thus, there are no enhancements from
the larger Dirac masses in these schemes, and also $R$ cannot be larger than 
in the hierarchical case because of other constraints.

Hence, it seems that an extension of the (3,~3) model is needed in order to address the LSND results.
Several authors~\cite{othermodels} have considered the introduction of Majorana masses of unknown origin
for the brane
states. In the
following, we propose two other extensions that seem to provide the necessary ingredients to accommodate the LSND
data. We give only simple estimates of the effects. To determine whether these
extensions can indeed provide a consistent explanation of all neutrino oscillation data requires a more careful
study, which is outside the scope of the present work.

\subsection{The (4,~4) Model}

Consider two extra sterile neutrinos, denoted by $N^s$ and
$\nu^s_L$, residing in the bulk and on the brane, respectively;
$\nu^s_L$ is left-handed. A mass term that couples these two
states is consistent with all the symmetries of the theory. It
has the form \beq {\cal L}_{mass} = \lambda \, M_F \,{\bar
\nu}^s_L \nu^s_R(x, 0) + {\rm h.c.}, \eeq{lmass} where $\lambda$
is a coupling analogous to $\lambda_{\alpha \beta}$ in
Eq.~\leqn{action}, and $\nu^s_R$ is the right-handed component of
$N^s$. Note that $M_F$ is the only natural choice for the mass
scale. Introducing a dimensionless Yukawa coupling $h=\lambda
M_F^{\delta/2}$ and performing the KK decomposition of $\nu^s_R$,
we obtain mass terms of the form \beq m_s \,\left(
\bar{\nu}_R^{s(0)} \nu_L^s + \sqrt{2}\sum_{n=1}^\infty
\bar{\nu}_R^{s(n)} \nu_L^s \right)\,+\,\,{\rm
h.c.}, \eeq{smassterms} where the
mass scale is given by \beq m_s = h {M_F^2 \over \Mp}. \eeq{smass}
For $M_F\sim 100$ TeV and $h \sim 1$, this mass scale is in the 1
to 10 eV range. Thus, the scale required to account for the LSND
results has been obtained without any fine-tuning, using only the
natural parameters of the theory. Note that the symmetries of the
theory also allow the brane Yukawa couplings between $\nu^s_R$
and the active left-handed neutrinos $\nu_L^\alpha$, which will
lead to mixings between the active species and the fourth sterile
KK tower. It seems plausible that the solar, atmospheric and LSND
results could be accommodated simultaneously in this framework;
however, a more careful analysis is clearly called for to verify
this.

\if
We denote the new 4-dimensional mass scale for the sterile-sterile coupling by $m^s$.  Then,
comparing the mass term in Eq.~\leqn{action} with the one in Eq.~\leqn{lmass} we obtain\beq \frac{m^s}{m^D_{\alpha
\beta}} = (h/h_{\alpha \beta}) \frac{M_F}{v}, \eeq{mratio} where the dimensionless constant $h$ is analogous to
$h_{\alpha \beta}$ defined in Eq.~\leqn{yukawa}.

For $M_F \sim 100$ TeV and $v = 246$ GeV, we find $m^s/m^D_{\alpha \beta} \sim 400 (h/h_{\alpha \beta})$.  Hence, for
$(h/h_{\alpha \beta}) \lsim 1$, $m^s$ can be in the 1 to 10 eV  range.  This new mass will set the scale for the
zero mode mass of the $4^{th}$ state, and therefore can yield a $3+1$ scheme.
Allowing analogous couplings to all of the bulk states, the mixing matrix elements
$l^{\alpha 4}$ could in general be large, although the consistency of this assumption needs further study.  Therefore,
we may expect that this (4,~4) model could in principle accommodate the LSND results.  As mentioned before, an
interesting feature of this proposal is that it does not require introducing {\it ad hoc} parameters and utilizes only
the natural parameters of the LED setup\footnote{One could, however, generate $2+2$ models with arbitary mixings by
allowing very small couplings analogous to $h$.}. In contrast with most other four neutrino schemes, there
are no Majorana mass terms and there is therefore a conserved lepton number.
\fi

\subsection{Two Large Dimensions}

As discussed above, the small rate expected for LSND in the
canonical (3,3) model is due to a cancellation between the
contributions of two KK towers, so that even in the inverted and
degenerate schemes the rate is proportional to $|\xi_2^2 -
\xi^2_1|^2$ or $|l^{e 3}|^2 |\xi_3^2 - \xi^2_2|^2$ rather than
$|\xi_2|^4$ or $|l^{e 3}|^2 |\xi_3|^4$. This cancellation can be
traced to the fact that all of the bulk neutrinos propagate in
the same large dimension. The cancellation could be broken in the
case in which there are two or more extra dimensions of unequal
radii. In the following, we consider a simple phenomenological
model  in which the three bulk neutrinos each propagate in only
one extra dimension, of radii $R_1$, $R_2$, and $R_3$,
respectively, where two or more of the $R_i$ may be different. \if
It is possible that a given bulk neutrino for some reason does not
propagate in the other dimensions, or is in a zero mode with
respect to the other dimensions. \fi We make no attempt to
construct a full model or derive all of the constraints, for our
goal is only to illustrate a possibility.

One can then show  that in the limit when the $1/R_i$ are all
large compared to the Dirac masses $m^D_i$, the amplitude $A_{\mu
e}(L)$ in \leqn{probab} becomes \beq A_{\mu e}(L) = \left[
-\frac{\pi^2}{6} \Xi \Xi^\dagger + \Xi P\Xi^\dagger \right]_{\mu
e} + \ldots, \eeq{unequal1} where $\Xi \equiv \sqrt{2} m^D {\bf
R}$, $m^D$ is the analog of the Dirac mass matrix in
\leqn{massmatrix}, ${\bf R}$ is the $3 \times 3$ matrix
${\rm\,diag\,}(R_1, R_2, R_3)$, and the matrix $P$ is given by
\beq P \equiv \sum_{k=1}^\infty \frac{e^{i \phi^k}}{k^2},
\eeq{pdef} with $ \phi^k \equiv [(k^2 L)/(2 E_\nu)] {\bf
R^{-2}}$. Writing $m^D= l^\dagger m_D^d r,$ where $m_D^d =
{\rm\,diag\,}(m^D_1, m^D_2, m^D_3)$ is diagonal, we see that
$A_{\mu e}(L)$ does not vanish for $m^D_1 \approx m^D_2$ provided
that the left and right unitary matrices $l$ and $r$ are not
equal. As a simple two-family example with $l$ and $r$ given by
rotations with angles $\beta_L$ and $\beta_R$, respectively, one
finds \beq A_{\mu e}(L) = 2 (m^{D}_2)^2 \left( -s_L c_R  + c_L
s_R\right) \left( c_L c_R + s_L s_R \right) \left[ \left( P_{11} -
\frac{\pi^2}{6} \right) R^2_1 - \left( P_{22} - \frac{\pi^2}{6}
\right) R^2_2 \right], \eeq{unequal2} where $s_{L,R} \equiv \sin
\beta_{L,R}$, and similarly for $c_{L,R}$. This is easily
generalized to the three-family case that can accommodate
atmospheric neutrino results. Thus, it is clear that the $\nu_e
\tofro \nu_\mu$ oscillation rates can be enhanced compared to the
(3,~3) model. However, a full investigation of this proposal is
beyond the scope of this paper.

\section{Conclusions}
\label{sec:conc}

In this paper, we have studied neutrino oscillations in the
context of theories with LED's.  In particular, we focused on a
5-dimensional model with three active brane and three sterile
bulk neutrinos, coupled to the 4-dimensional Higgs field. This
setup yields small Dirac masses for the active species, in
addition to a tower of sterile KK neutrinos.  An attractive
feature of this (3,~3) model is that it does not introduce any
{\it ad hoc} parameters, such as small 4-dimensional Majorana
masses.  With an extra dimension of radius $R$, the lightest KK
neutrinos have masses $\sim 1/R$. Our approach was based on the
assumption that a three-flavor analysis of the data provides the
best fit for the observed oscillations, and we have treated the
effect of the LED as a perturbation. Hence, in our treatment, the
Dirac masses are taken to be much smaller than $1/R$.

The recent solar neutrino data from the SNO collaboration, in
conjunction with the results of the Super-Kamiokande experiment
for atmospheric neutrinos, significantly constrain the size of
the effect of sterile states on the solar neutrino deficit.  In
particular, models based solely on active-sterile oscillations,
as in Ref.~\cite{DS}, are now ruled out. This fact motivated us
to determine what bounds the new data place on the contributions
of KK sterile states to neutrino oscillations. Since these
sterile states originate in the extra dimensions, we obtained
constraints on the size $R$ of the largest of the extra
dimensions. In addition to the new solar data analysis, we also
performed analyses of reactor, accelerator, and atmospheric
data.  The (3,~3) model studied here cannot accommodate the LSND
data. However, we proposed two natural extensions of this model
to address the LSND results. One is a (4,~4) model with two extra
sterile neutrinos, one on the brane and another in the bulk,
which can yield a $3+1$, or, less naturally, a $2+2$
four-neutrino scheme. These are phenomenologically similar to
four neutrino schemes obtained in other theoretical frameworks,
yielding equally good or poor descriptions of the data.  However,
there are no Majorana masses, so there is a conserved lepton
number  and neutrinoless double beta decay is forbidden. The
other extension uses two LED's of unequal radii, in each of which
 a bulk neutrino is sequestered.  We provided estimates of
the effects in these two cases and concluded that the LSND results seem to be accommodated in our proposals, although a
more careful analysis is warranted.

In this paper, we have concentrated on constraints from laboratory, solar and atmospheric
neutrino oscillation
experiments. There are also complementary constraints from cosmology and astrophysics.
In particular, the degenerate scheme may be excluded by considerations of big bang nucleosynthesis, cosmic microwave
background radiation, and diffuse extra-galactic background radiation~\cite{cosmology}, but
the hierarchical and
inverted schemes are consistent. There are also potentially very strong 
limits on all of the
schemes coming from the constraints on the energy carried away by the 
sterile KK modes from deep inside a supernova core.
Qualitative estimates~\cite{SNbound} concentrating on a single family suggest
that the constraints may be considerably stronger than those in Table
1. A more detailed study would be very useful. Bounds on the model considered 
here can also be derived from studying the effects of KK neutrinos in 
electroweak processes~\cite{Apostolos}.

Our main results are contained in Table~\ref{tab:rabounds}. These are to be interpreted
as constraints on the size of the largest of the extra dimensions, regardless of their
total number.
If, as seems plausible, the
pattern of masses in the neutrino sector is like those of the other fermionic sectors,
and hence hierarchical, the strongest bound comes from the atmospheric neutrino data
and requires $R$ to be less than about $0.8 \, \mu$m. The bounds in the inverted and
degenerate mass schemes are even stronger. On the other hand, the currently discussed
Cavendish type experiments have sensitivities only of order 50 $\mu$m, and an
improvement of two orders of magnitude would be necessary to reach the region allowed
in our model. 
Thus, to the degree that the (3,~3) model is a natural context for small
Dirac masses in theories with LED's, our analysis suggests that neutrino 
oscillations seem to
rule out the possibility of observing the gravitational effects of the
extra dimensions in the foreseeable future.

\section{Acknowlegements}
It is a pleasure to thank J. Bahcall, T. Kajita, C. Lunardini, and
A. Pierce for useful communications and discussions.  H. D. is
supported in part by the Department of Energy, under grant
DE-FG02-90ER40542.  One of us (P. L.) was  supported by the W. M.
Keck Foundation as a Keck Distinguished Visiting Professor at the
Institute for Advanced Study and by the Monell Foundation, and by
the Department of Energy  grant DOE-EY-76-02-3071.  M. P. is
supported by the Director, Office of Science, Office of High
Energy and Nuclear Physics, of the U. S. Department of Energy
under Contract DE-AC03-76SF00098.

\end{document}